\documentclass{article}
\usepackage{PRIMEarxiv}

\usepackage[utf8]{inputenc} 
\usepackage[T1]{fontenc}    
\usepackage{hyperref}       
\usepackage{url}            
\usepackage{booktabs}       
\usepackage{amsfonts}       
\usepackage{nicefrac}       
\usepackage{microtype}      
\usepackage{lipsum}
\usepackage{fancyhdr}       
\usepackage{graphicx}       
\usepackage{tabularx}
\graphicspath{{media/}}     
\usepackage{varwidth}
\usepackage{adjustbox}
\usepackage{makecell}
\usepackage[justification=centering]{caption}

\usepackage{booktabs, multirow} 
\usepackage{soul}
\usepackage[table]{xcolor} 
\usepackage{changepage,threeparttable} 

\title{
Assessing Code Generation with Intermediate Languages

}

\author{
  Xun Deng\thanks{Equal Contribution.}~~,~~Sicheng Zhong$^{*}$ \\
  University of Toronto \\
  \texttt{\{xun.deng, sicheng.zhong\}@mail.utoronto.ca} \\
   \And
  Honghua Dong \\
  University of Toronto \& Vector Institute\\
  \texttt{honghuad@cs.toronto.edu} \\
  \And
  Jingyu Hu \\
  University of Toronto \\
  \texttt{jingyu.hu@mail.utoronto.ca} \\
  \And
  Sidi Mohamed Beillahi \\
  University of Toronto \\
  \texttt{sm.beillahi@utoronto.ca} \\
  \And
  Xujie Si \\
  University of Toronto \& Vector Institute\\
  \texttt{six@cs.toronto.edu} \\
  \And
  Fan Long \\
  University of Toronto \\
  \texttt{fanl@cs.toronto.edu} \\
}

\begin{document}
\maketitle
\begin{abstract}
Intermediate step methodologies like chain of thoughts (COT) have demonstrated effectiveness in enhancing the performance of Large Language Models (LLMs) on code generation. This study explores the utilization of intermediate languages, including various programming languages, natural language solutions, and pseudo-code, and systematically evaluates their impact on the performance of LLMs in code generation tasks. Our experiments encompass eleven models across the CodeLlama, GPT, and Mistral families, as well as newly released smaller models. Our findings reveal that intermediate languages generally exhibit greater efficacy in larger models that have not yet achieved state-of-the-art performance. Natural language consistently emerges as the most effective intermediate representation across all target languages. However, we observe no universally effective intermediate formal language across different models and target languages. Furthermore, we uncover a weak correlation between the correctness of intermediate solutions and final generation, suggesting that improvements may stem from the chain-of-thought effect rather than language-specific transfer. Interestingly, we discover that for GPT family models, prompting multiple times without explicit self-correction instructions yields performance gains across the studied languages.
\end{abstract}

\section{Introduction} 

Large Language Models (LLMs) have demonstrated significant potential in assisting with code generation, a key challenge in programming languages and software engineering. With a programming task description in natural language, LLM-powered tools can automatically synthesize programs in a target language within seconds~\cite{brown2020language}. By leveraging their extensive training data, these tools frequently outperform traditional program synthesis techniques~\cite{Alur_2017, gulwani2011automating}, completing complex code generation tasks with enhanced efficiency and accuracy. However, despite recent advancements in LLM capabilities, a notable gap remains between AI-driven code generation tools and human developers in performing coding tasks~\cite{jimenez2024swebench}.

One lightweight and effective approach to enhancing LLM-powered tools is in-context learning (a.k.a prompt engineering), which provides appropriate guidance or demonstrations along with the original problem description in the prompt. 
Prior research has shown that guiding LLMs through intermediate steps can significantly improve final outcomes across various tasks. Simple guidance, such as including phrases like "Let's think step-by-step" in the prompt, can enhance performance~\cite{wei2022chain}. Interestingly, it has been found that code generation can serve as an intermediate step for other tasks. Researchers have discovered that introducing domain-specific languages and asking LLMs to generate code in these intermediate languages can improve LLM performance in math problem-solving and visual recognition tasks~\cite{nye2021work}.

This paper investigates whether introducing intermediate steps and intermediate languages can enhance LLM code generation. We designed prompts to guide LLMs to outline an intermediate solution in pseudo-code, natural language, or code in a different language before generating the final code in the target language. We compared the code generated using our prompts with direct generation on the HumanEval benchmark~\cite{humanevalx}. To obtain comprehensive results, we studied five different programming languages, Python, Java, Cpp, Go, and Rust, and a range of LLM families, including CodeLlama, GPT, and Mixtral, as well as three newly released smaller models.


Our results demonstrate that the proposed intermediate-step methods can generate better code across different models and all five benchmark languages. We observed more consistent improvements in larger models with over 10 billion parameters and in scenarios where direct generation did not achieve very high accuracy (i.e., less than 25\%). Notably, among all intermediate languages, natural language performed the best on average across all large models.

To understand the reasons behind the performance improvement, we investigated the correlations between the correctness of intermediate results and the final code generation results. Surprisingly, we found that this correlation was not substantial and often did not exceed the baseline correlation of the benchmark set. One possible implication is that the improvement may stem from the chain-of-thought effect, i.e., by encouraging LLMs to generate additional intermediate information (even if not fully correct), our prompt method allows LLMs to operate within a more productive context, leading to better final results.

Additionally, we observed another interesting phenomenon that validates the existence of the chain-of-thought effect in LLM code generation. Prompting the model multiple times without explicitly instructing it to self-correct resulted in performance gains in many scenarios, particularly within the GPT family. Although the self-correcting ability of LLMs is still under debate~\cite{huang2024largelanguagemodelsselfcorrect}, these findings highlight the complexity of LLM behavior and underscore the need for further investigation into the mechanisms that make intermediate answers effective in guiding the model's reasoning process.

This paper makes the following contributions: 1) We present the first systematic study on utilizing intermediate languages in code generation via LLMs; 2) We demonstrate that code generation through intermediate languages, on average, outperforms direct generation across large models and various target programming languages; 3) We identify a weak correlation between the correctness of intermediate results and final code correctness, proposing the chain-of-thought effect as a possible explanation for the observed performance improvement.

The paper is organized as follows: We give an example of our prompting method
in Section~\ref{sec:overview}. We present our experiment design and set up in
Section~\ref{sec:design} explain the experiment design and set up and present
the experiment result and discussion in Section~\ref{sec:experiment}. We
discuss related work in Section~\ref{sec:relatedwork} and conclude in
Section~\ref{sec:conclusion}.

\section{Example and Overview}\label{sec:overview}


\begin{figure}[!htb]
\includegraphics[width=\textwidth]{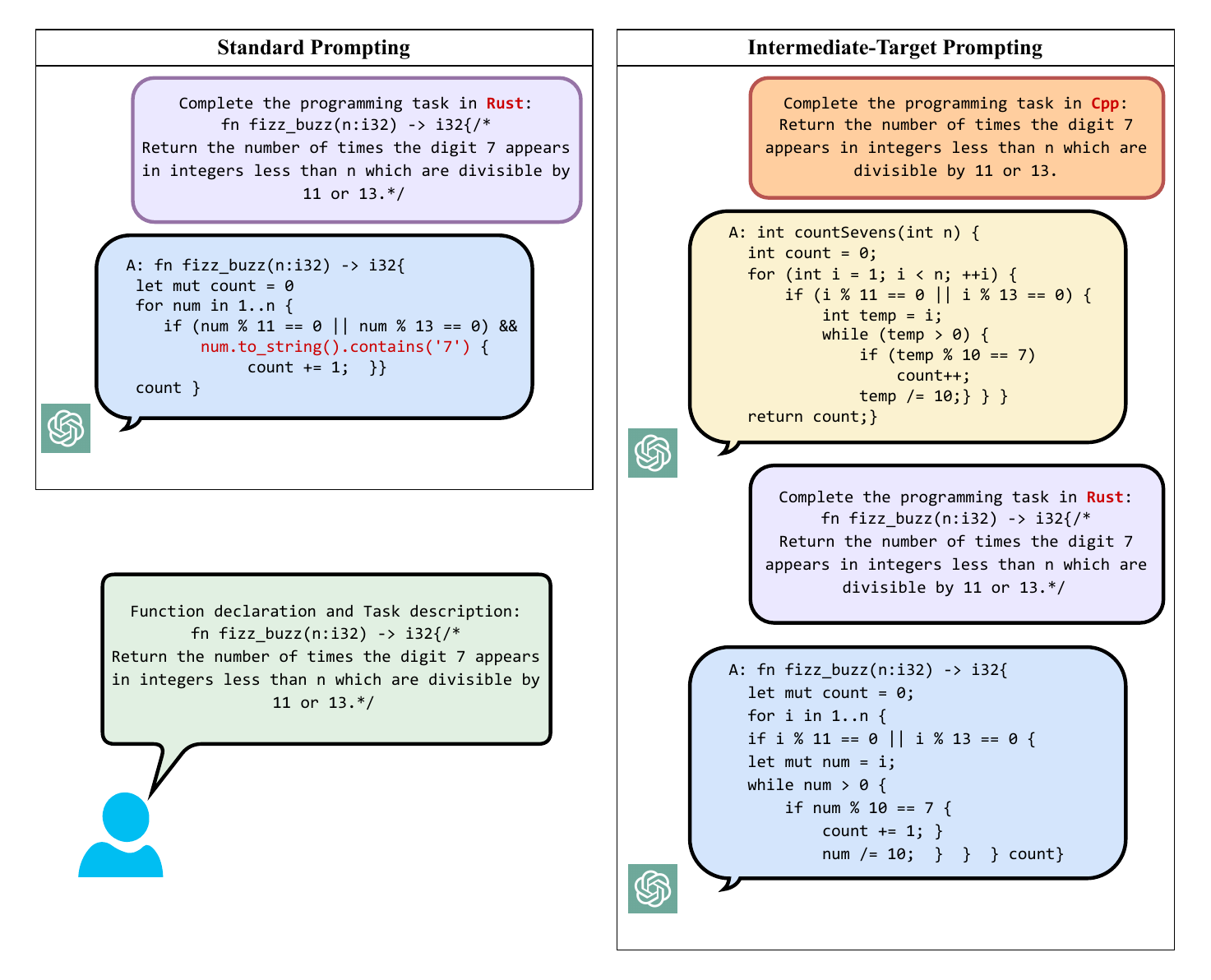}
\caption{Comparison between the flow of standard prompting and the flow of intermediate-target prompting method. In the intermediate-target prompting, we first prompt the model to generate code in an intermediate language with the task description only. Then, we prompt the model to generate a solution in the target language.}\label{fig:prompt}
\end{figure}

In this section, we introduce the concept of intermediate-step prompting. Our proposed prompting method involves a two-stage approach: the user gives a prompt containing a task description and function declaration, and instead of directly prompting the LLM for a solution in the target language, we first prompt the model to generate a solution in an intermediate language. To illustrate this method, we present a motivation from HumanEval-X benchmark~\cite{humanevalx}. Specifically, the programming task is to return the number of times the digit 7 appears in less than \textit{n} divisible by 11 or 13, with the target code generation language being Rust. The flow of standard prompting is shown on the right side of Figure~\ref{fig:prompt}. Although the output code is syntactically correct, it contains logic errors. Specifically, the model incorrectly uses the built-in function \textit{contains}, which checks if the number includes the digit seven rather than counting the total occurrences of the digit seven as the prompt requires.

In this example, we use Cpp as the intermediate language. At this stage, we only provide the task description, as some models fail to generate a solution in the intermediate language when the function declaration in the target language is also provided. Then, we prompt the model again, given both the task description and function declaration and request a solution in target language. The model gives a logic correct code in the Cpp-intermediate step and with this context, the final rust generation is correct. 

As humans tend to write a solution sketch or pseudo-code when solving reasoning and programming tasks, we also evaluate the model's code generation ability using informal languages, such as natural language or pseudo-code, as the intermediate step. This evaluation aims to determine if there is a universal intermediate representation that can enhance code generation for all target languages. To understand the observed improvements using the proposed method and to assess whether the correctness of the intermediate solution is an influencing factor, we enforce the intermediate generation to be ground truth and analyze the correlations between direct generations and intermediate-step prompting.

In the proposed prompting flow, the model has two chances to complete the same task. In contrast, standard prompting provides the model with a one-shot opportunity. Therefore, we explore the model's behavior under repetitive prompting, \emph{i.e.}, prompting the model to complete the same task in the same languages multiple times. We now present our experiment design details.

\section{Experiment Design and Set up}\label{sec:design}
\subsection{Intermediate Languages Selection}
We deliberately select five programming languages: Cpp, Go, Java, Python, and Rust, representing distinct paradigms with diverse levels of popularity within the global developer community. For informal languages, we consider natural language solution sketch and pseudo code.

\subsection{Models}
We evaluate 11 models. From the CodeLlama family, we assess 7b-Instruct, 13b-Instruct, 34b-Instruct, and 70b-Instruct~\cite{rozière2024code}. We also evaluate GPT3.5 (gpt-3.5-turbo-1106) and GPT4 (gpt-4-1106-preview)~\cite{brown2020language} 
We select two models from the Mistral family, namely Mistral-7b-Instruct and Mixtral-8x7b-Instruct~\cite{jiang2023mistral, jiang2024mixtral}. Additionally, we perform experiments on Deepseek-coder-instruct-1.3b, Stable-code-instruct-3b and Phi-3-mini-128k-instruct~\cite{guo2024deepseekcoder, pinnaparaju2024stable, abdin2024phi3}. Table~\ref{tab:models} summarizes the key features of these models, including model size, MMLU score (where applicable), architecture, context window, modality, and whether they are code-based.

\begin{table}[!tb]\centering
\caption{Summary of features for various language models.}\label{tab:models}
\begin{tabular}{l|ccccccc}\toprule
&\makecell{Model\\size}&\makecell{MMLU\\score} &\makecell{Context\\window} &Modality &Code-focused \\\midrule
Mistral 7b &7.3b &70.60\% &128k &pretrained &No \\
Mixtral 8x7b (MOE) &48b &60.10\% &32k &pretrained &No \\
\midrule
CodeLlama7b-instruct &7b &N/A &16k &SFT &Yes \\
CodeLlama13b-instruct &13b &N/A &16k &SFT &Yes \\
CodeLlama34b-instruct &34b &N/A &16k &SFT &Yes \\
CodeLlama70b-instruct &70b &N/A &16k &SFT&Yes \\
\midrule
GPT35 &unknown &70.00\% &16k &SFT&No \\
GPT4 &unknown&86.40\% &128k &SFT &No \\
\midrule
Stable-code-instruct-3b &2.7b &N/A  &16k &SFT &Yes \\
Deepseek-coder-1.3b-instruct &1.3b &N/A &16k &SFT&Yes \\
Phi-3-mini-128k-instruct &3.8b &68.10\%  &128k &SFT &No \\
\bottomrule
\end{tabular}
\vspace{-1em}
\end{table}


\subsection{Benchmarks}
We choose to use the HumanEval-X benchmark, which is the multilingual version of the standard HumanEval benchmark~\cite{humanevalx, chen2021evaluatinglargelanguagemodels}. The benchmark consists of 164 programming questions covering natural language comprehension, algorithms, and simple math problems. 

\subsection{Experiment design}
Our study comprises three experiments, designed to comprehensively evaluate the effectiveness of intermediate-step prompting, the impact of intermediate solutions, and the models' self-correction abilities in code generation tasks.

\textbf{Intermediate Language Prompting:} Firstly, we follow the methodology described in Section~\ref{sec:overview} and use a specific programming language as an intermediary. We explore all the intermediate-target combinations for the five selected languages. To evaluate the effectiveness of using natural language and pseudo-code, we first prompt the model to generate a solution in natural language or pseudo-code and then generate it in the target language. 

\textbf{Ground Truth Intermediate Solution:} Instead of prompting the model to generate an intermediate solution, we provide the correct solution (ground truth)in the intermediate language and then prompt the model to generate in the target language. This evaluates how the intermediate solution's correctness affects the target generation's quality.

\textbf{Self-Correction Capability:} We assess the model's ability to self-correct by providing the generation history and repeatedly prompting the model \textit{k} times. Importantly, we do not provide any feedback or correction instructions in the prompt, allowing us to observe the model's autonomous improvement capabilities.

\section{Experiment Results and Discussions}\label{sec:experiment}
In this section, we present the experiment results and answer the following research questions.\\
\textbf{RQ1:} Does generate code in another programming language as an intermediate step improve the performance? \\
\textbf{RQ2:} How does pseudo code and solution sketch in natural language affect the model's performance? \\
\textbf{RQ3:} Are intermediate language and target language generations correlated?\\
\textbf{RQ4:} How does the number of repeat prompting affect the model's performance without providing instructions such as self-correcting? \\

\subsection{Effect of intermediate representation on LLMs}
In this section, we present the experiment results for the selected large language models and answer the first two RQs.
\subsubsection{CodeLlama Family}
We present the experimental results for the CodeLlama Family in Tables~\ref{tab:codellama}. The first column indicates the intermediate languages. Specifically, for codeLlama-7b, the most significant performance gain of 4.3\% is observed in Go generation using Cpp as an intermediate language. Except for Go, direct generations either achieve the highest pass rate or are comparable to the best performance. The use of natural language solution sketches or pseudo code yields similar performance for this model. Notably, performance degradation after utilizing intermediate steps is more pronounced in C++, Java, and Python compared to Go and Rust.

For the 13b model, Rust emerges as the most effective intermediate programming language. We observe approximately a 5\% performance gain in Java and Python generation when using Rust as an intermediary. Interestingly, while Rust shows positive effects in generating other languages, minimal improvement is observed when it's the target language, possibly due to the inherent difficulty of Rust generation for this model. For instance, Go and pseudo-code aid in performance improvement across the board, except for Rust. Conversely, while natural language doesn't yield favorable results in Go, Java, and Python generation, it enhances Rust generation.

In the case of Llama-34b, the most significant performance gains are observed in Rust and Java generation. However, it's worth noting that this could be attributed to the initially low accuracy in direct generation for Rust. Similar to the finding in Llama-13b, Rust emerges as the optimal intermediate language for this model, enhancing generations across the other four languages. Interestingly, the utilization of natural language or pseudo code yields the most substantial improvements in Java generation, whereas little to no performance gain is observed in other languages. Furthermore, this model benefits more from the proposed intermediate-target prompting compared to its 13b and 7b counterparts. This suggests that the effectiveness of intermediate-step prompting may scale with model size within the CodeLlama family.

For CodeLlama70b, the state-of-art of the CodeLlama family, we note a reduced performance gain with intermediate languages compared to the smaller models, particularly evident in the generation of Java, Python, and Rust. While utilizing Rust yields a 3.7\% and 2.4\% performance improvement compared to direct generation, direct generation results in comparable outcomes to the highest accuracy achieved. Overall, employing intermediate languages does not lead to significant improvement or deterioration compared to direct generation. The sole exception is when using Python as an intermediate language for Rust generation, where a performance decrease of 12\% is observed. The utilization of natural language solution sketches and pseudo code results in a similar generation success rate. 

\textbf{Key findings}:
Prior to achieving the state-of-the-art performance represented by CodeLlama-70b, we observed that the utilization of intermediate languages has a more pronounced positive effect on code generation for larger models. Notably, in the tables, a higher number of cells are marked green for models 13b and 34b compared to 7b.

\begin{table}[!htp]\centering
\caption{Experiment results of CodeLlama family.
Before achieving the state-of-the-art performance demonstrated by CodeLlama-70b, using intermediate languages enhances code generation in larger models.}\label{tab:codellama}
\begin{adjustbox}{width=\linewidth}
\begin{tabular}{l|ccccc|ccccc}\toprule
&\multicolumn{5}{c|}{CodeLlama7b-instruct} &\multicolumn{5}{c}{CodeLlama13b-instruct} \\\cmidrule{2-6}\cmidrule{7-11}
&Cpp &Go &Java &Python &Rust &Cpp &Go &Java &Python &Rust \\\midrule
Cpp &- &\cellcolor[HTML]{d9ead3}4.3\% &-3.7\% &-3.7\% &-4.9\% &- &\cellcolor[HTML]{d9ead3}0.6\% &-2.4\% &\cellcolor[HTML]{d9ead3}1.2\% &\cellcolor[HTML]{d9ead3}0.0\% \\
Go &-4.9\% &- &\cellcolor[HTML]{d9ead3}0.6\% &-2.4\% &-3.7\% &\cellcolor[HTML]{d9ead3}1.5\% &- &\cellcolor[HTML]{d9ead3}2.4\% &\cellcolor[HTML]{d9ead3}3.7\% &-3.7\% \\
Java &-3.0\% &-0.6\% &- &\cellcolor[HTML]{d9ead3}1.2\% &-1.8\% &-5.2\% &-1.8\% &- &-3.7\% &-3.0\% \\
Python &-3.7\% &-4.9\% &-3.0\% &- &-6.7\% &\cellcolor[HTML]{d9ead3}0.3\% &-2.4\% &\cellcolor[HTML]{d9ead3}4.9\% &- &-1.8\% \\
Rust &-3.0\% &0.0\% &-1.8\% &-1.2\% &- &\cellcolor[HTML]{d9ead3}0.3\% &\cellcolor[HTML]{d9ead3}0.6\% &\cellcolor[HTML]{d9ead3}5.5\% &\cellcolor[HTML]{d9ead3}4.9\% &- \\
Natural language &-3.7\% &-0.6\% &-6.7\% &-7.3\% &-1.8\% &\cellcolor[HTML]{d9ead3}0.3\% &-1.2\% &-2.4\% &-2.4\% &\cellcolor[HTML]{d9ead3}0.6\% \\
Pseudo code &-4.3\% &-1.2\% &-4.3\% &-4.9\% &-4.9\% &\cellcolor[HTML]{d9ead3}0.3\% &\cellcolor[HTML]{d9ead3}0.6\% &\cellcolor[HTML]{d9ead3}0.0\% &\cellcolor[HTML]{d9ead3}1.8\% &-4.9\% \\
Direct generation &\cellcolor[HTML]{fff2cc}32.3\% &\cellcolor[HTML]{fff2cc}26.8\% &\cellcolor[HTML]{fff2cc}37.8\% &\cellcolor[HTML]{fff2cc}39.6\% &\cellcolor[HTML]{fff2cc}23.8\% &\cellcolor[HTML]{fff2cc}33.8\% &\cellcolor[HTML]{fff2cc}26.8\% &\cellcolor[HTML]{fff2cc}37.8\% &\cellcolor[HTML]{fff2cc}39.6\% &\cellcolor[HTML]{fff2cc}25.0\% \\
\bottomrule
\multicolumn{11}{c}{}\\
\toprule
&\multicolumn{5}{c|}{CodeLlama34b-instruct} &\multicolumn{5}{c}{CodeLlama70b-instruct} \\\cmidrule{2-6}\cmidrule{7-11}
&Cpp &Go &Java &Python &Rust &Cpp &Go &Java &Python &Rust \\\midrule
Go &0.0\% &- &\cellcolor[HTML]{d9ead3}2.4\% &0.0\% &\cellcolor[HTML]{d9ead3}7.9\% &-0.6\% &- &-2.4\% &-3.0\% &-3.0\% \\
Java &-1.8\% &\cellcolor[HTML]{d9ead3}2.4\% &- &\cellcolor[HTML]{d9ead3}0.6\% &\cellcolor[HTML]{d9ead3}6.1\% &\cellcolor[HTML]{d9ead3}2.4\% &\cellcolor[HTML]{d9ead3}0.6\% &- &-4.3\% &-5.5\% \\
Python &-1.2\% &\cellcolor[HTML]{d9ead3}1.2\% &\cellcolor[HTML]{d9ead3}6.7\% &- &\cellcolor[HTML]{d9ead3}6.7\% &\cellcolor[HTML]{d9ead3}2.4\% &0.0\% &-3.0\% &- &-12.2\% \\
Rust &\cellcolor[HTML]{d9ead3}0.6\% &\cellcolor[HTML]{d9ead3}3.0\% &\cellcolor[HTML]{d9ead3}4.3\% &\cellcolor[HTML]{d9ead3}1.8\% &- &\cellcolor[HTML]{d9ead3}3.7\% &\cellcolor[HTML]{d9ead3}2.4\% &-0.6\% &-4.3\% &- \\
Natural language &\cellcolor[HTML]{d9ead3}0.6\% &-1.2\% &\cellcolor[HTML]{d9ead3}5.5\% &\cellcolor[HTML]{d9ead3}1.8\% &\cellcolor[HTML]{d9ead3}2.4\% &\cellcolor[HTML]{d9ead3}3.0\% &\cellcolor[HTML]{d9ead3}1.2\% &0.0\% &-3.0\% &-2.4\% \\
Pseudo code &-1.8\% &-1.2\% &\cellcolor[HTML]{d9ead3}4.3\% &-6.7\% &\cellcolor[HTML]{d9ead3}4.3\% &\cellcolor[HTML]{d9ead3}0.6\% &\cellcolor[HTML]{d9ead3}1.8\% &-0.6\% &\cellcolor[HTML]{d9ead3}0.6\% &-4.3\% \\
Direct generation &\cellcolor[HTML]{fff2cc}38.4\% &\cellcolor[HTML]{fff2cc}31.1\% &\cellcolor[HTML]{fff2cc}42.7\% &\cellcolor[HTML]{fff2cc}45.1\% &\cellcolor[HTML]{fff2cc}20.7\% &\cellcolor[HTML]{fff2cc}52.4\% &\cellcolor[HTML]{fff2cc}50.6\% &\cellcolor[HTML]{fff2cc}65.9\% &\cellcolor[HTML]{fff2cc}68.3\% &\cellcolor[HTML]{fff2cc}32.9\% \\
\bottomrule
\end{tabular}

\end{adjustbox}
\end{table}

\begin{table}[!htp]\centering
\caption{Experiment Results of GPT family. Natural language serves as a more effective intermediate representation than programming languages.}\label{tab:gpts}
\begin{adjustbox}{width=\linewidth}
\begin{tabular}{l|ccccc|ccccc}\toprule
&\multicolumn{5}{c|}{GPT35} &\multicolumn{5}{c}{GPT4} \\\cmidrule{2-6}\cmidrule{7-11}
&Cpp &Go &Java &Python &Rust &Cpp &Go &Java &Python &Rust \\\midrule
Cpp &- &-8.2\% &\cellcolor[HTML]{d9ead3}5.5\% &-7.7\% &\cellcolor[HTML]{d9ead3}0.1\% &- &-9.8\% &\cellcolor[HTML]{d9ead3}1.2\% &0.0\% &-6.1\% \\
Go &-4.1\% &- &\cellcolor[HTML]{d9ead3}6.8\% &-8.7\% &-0.6\% &-11.6\% &- &-1.2\% &-1.8\% &-9.8\% \\
Java &-2.9\% &-2.9\% &- &-6.3\% &-1.3\% &-9.1\% &-11.6\% &- &\cellcolor[HTML]{d9ead3}3.0\% &-4.9\% \\
Python &-1.6\% &-0.2\% &-0.5\% &- &-1.7\% &-5.5\% &-18.3\% &-6.1\% &- &-4.3\% \\
Rust &-5.1\% &-15.4\% &\cellcolor[HTML]{d9ead3}3.9\% &-11.3\% &- &-8.5\% &-12.2\% &-3.7\% &0.0\% &- \\
Natural language &\cellcolor[HTML]{d9ead3}0.4\% &\cellcolor[HTML]{d9ead3}2.6\% &\cellcolor[HTML]{d9ead3}8.2\% &-4.6\% &\cellcolor[HTML]{d9ead3}1.6\% &\cellcolor[HTML]{d9ead3}3.0\% &-2.4\% &\cellcolor[HTML]{d9ead3}6.1\% &\cellcolor[HTML]{d9ead3}6.1\% &-0.6\% \\
Pseudo code &-12.4\% &-4.1\% &\cellcolor[HTML]{d9ead3}8.8\% &-8.3\% &-2.1\% &-13.4\% &-6.1\% &\cellcolor[HTML]{d9ead3}6.7\% &\cellcolor[HTML]{d9ead3}3.0\% &-14.6\% \\
Direct generation &\cellcolor[HTML]{fff2cc}50.9\% &\cellcolor[HTML]{fff2cc}57.2\% &\cellcolor[HTML]{fff2cc}54.0\% &\cellcolor[HTML]{fff2cc}70.5\% &\cellcolor[HTML]{fff2cc}44.8\% &\cellcolor[HTML]{fff2cc}61.0\% &\cellcolor[HTML]{fff2cc}68.3\% &\cellcolor[HTML]{fff2cc}67.7\% &\cellcolor[HTML]{fff2cc}76.8\% &\cellcolor[HTML]{fff2cc}42.1\% \\
\bottomrule
\end{tabular}
\end{adjustbox}
\end{table}

\subsubsection{GPT Family}
We present the experimental results for the GPT models in Table~\ref{tab:gpts}. In the case of GPT-3.5, we observe that generation for Java benefits from using intermediate languages (except Python), with the most significant performance gain observed when employing pseudo code as the intermediary, resulting in an approximately 9\% improvement. However, using other programming languages does not show a positive effect on the generation of Cpp, Go, and Python. Apart from Java, the only instance where another programming language provided assistance is in Rust generation with Cpp as an intermediary, albeit with only a slight improvement observed. Interestingly, natural language emerges as the most effective intermediate language for GPT-3.5. Pseudo code, except in the case of Java, does not offer notable benefits. Furthermore, significant performance degradation is observed in Cpp and Python when pseudo code is utilized for generation.

In the case of GPT-4, we observed that Java and Python benefited from the use of other programming languages such as Cpp. However, the most significant accuracy gains are observed when employing natural language and pseudo code as intermediate languages, resulting in improvements of 6.1\% and 6.7\% in Java and Python generation, respectively. Similar to GPT-3.5, natural language yields the best performance across all five languages, showcasing improvements or comparable results compared to direct generation. While notable improvements are observed when utilizing pseudo code in the generation of Java and Python, significant degradations are found in Cpp and Rust. 

\textbf{Key findings}:
For GPT family, natural language shows better performance as an intermediate representation than programming languages. Pseudo code results in performance drop in languages such as Cpp and Rust.

\subsubsection{Mistral Family}
We present the experimental results for the Mistral models in Table~\ref{tab:mis}. For Mistral 7b, using an intermediate language shows little improvement in C++, Go, Java, and Python generation. The model exhibits poor performance in Rust, both with and without intermediate generation, indicating limited knowledge of this language. Notably, we observe significant performance degradation in certain cases. For instance, Java generation experiences nearly a 10\% decrease when using C++ as an intermediate language. Similarly, Go and Python exhibit performance decreases of around 5\% when using Rust as an intermediary. In most cases, pseudo-code and natural language yield comparable results to direct generation. The exception is Python generation using pseudo-code as an intermediate step, where a 7\% drop is observed.

For the Mixtral 8x7b model, we observe significant performance gains for languages such as Go and Rust through intermediate generations. Using Java and Python as intermediate languages for Go improves performance by over 6\%. Similar improvement is observed in Rust generation when using natural language as an intermediate step, with a 4\% gain when using Go. Notably, using natural language as an intermediary increases Python generation accuracy from 51\% to 57\%. While pseudo-code generally improves performance for most languages, natural language demonstrates more substantial improvement for this particular model.

\textbf{Key findings}:
The positive effect of using intermediate representation is more significant on Mixtral-8x7b than the smaller 7b model. Languages in which the model demonstrates less knowledge benefit more from the proposed prompting method. 

\begin{table}[!htp]\centering
\caption{Experiment results of Mistral family. 
The positive impact of using intermediate representation is more pronounced on Mixtral-8x7b compared to the smaller 7b model. Languages in which the model shows less proficiency benefit more from the proposed prompting method.}\label{tab:mis}
\begin{adjustbox}{width = \linewidth}
\begin{tabular}{l|ccccc|ccccc}\toprule
&\multicolumn{5}{c|}{Mistral-7b} &\multicolumn{5}{c}{Mixtral-8x7b} \\\cmidrule{2-6}\cmidrule{7-11}
&Cpp &Go &Java &Python &Rust &Cpp &Go &Java &Python &Rust \\\midrule
Cpp &- &0.0\% &-8.9\% &-4.4\% &\cellcolor[HTML]{d9ead3}1.0\% &- &\cellcolor[HTML]{d9ead3}3.7\% &-7.9\% &-0.6\% &-0.6\% \\
Go &-4.0\% &- &-5.2\% &-4.8\% &-0.1\% &-6.7\% &- &-6.7\% &-8.5\% &\cellcolor[HTML]{d9ead3}4.3\% \\
Java &-3.7\% &-0.9\% &- &-4.3\% &\cellcolor[HTML]{d9ead3}1.1\% &-3.0\% &\cellcolor[HTML]{d9ead3}6.1\% &- &0.0\% &\cellcolor[HTML]{d9ead3}3.0\% \\
Python &-0.2\% &-2.1\% &-4.5\% &- &\cellcolor[HTML]{d9ead3}1.7\% &-4.3\% &\cellcolor[HTML]{d9ead3}6.1\% &-8.5\% &- &-5.5\% \\
Rust &-2.4\% &-4.1\% &-4.5\% &-5.4\% &- &-4.9\% &\cellcolor[HTML]{d9ead3}4.3\% &-9.1\% &\cellcolor[HTML]{d9ead3}1.2\% &- \\
Natural language &-0.5\% &\cellcolor[HTML]{d9ead3}1.1\% &-3.5\% &-0.5\% &-0.6\% &\cellcolor[HTML]{d9ead3}1.2\% &\cellcolor[HTML]{d9ead3}3.7\% &-1.2\% &\cellcolor[HTML]{d9ead3}6.1\% &\cellcolor[HTML]{d9ead3}6.1\% \\
Pseudo code &-3.4\% &-3.3\% &-2.4\% &-7.1\% &0.0\% &-5.5\% &0.0\% &-7.9\% &\cellcolor[HTML]{d9ead3}0.6\% &\cellcolor[HTML]{d9ead3}1.2\% \\
Direct generation &\cellcolor[HTML]{fff2cc}22.9\% &\cellcolor[HTML]{fff2cc}19.4\% &\cellcolor[HTML]{fff2cc}29.6\% &\cellcolor[HTML]{fff2cc}31.8\% &\cellcolor[HTML]{fff2cc}7.2\% &\cellcolor[HTML]{fff2cc}42.1\% &\cellcolor[HTML]{fff2cc}40.2\% &\cellcolor[HTML]{fff2cc}56.1\% &\cellcolor[HTML]{fff2cc}51.2\% &\cellcolor[HTML]{fff2cc}29.9\% \\
\bottomrule
\end{tabular} 
\end{adjustbox}
\end{table}

\begin{table}[!htb]
\centering
\caption{Experiment results of Deepseek-1.3b,  Stable-3b, and Phi-3-mini-128k-instruct. The positive impact of using an intermediate language is not significant. The intermediate step is most beneficial for languages in which the model has limited knowledge.}\label{tab:small3}
\begin{adjustbox}{width = \linewidth}
\begin{tabular}{l|ccccc|ccccc}\toprule
&\multicolumn{5}{c}{Stable-code-3b} &\multicolumn{5}{c}{Deepseek-code-1.3b} \\\cmidrule{2-11}
&Cpp &Go &Java &Python &Rust &Cpp &Go &Java &Python &Rust \\\midrule
Cpp &- &-8.5\% &-9.1\% &-11.0\% &0.0\% &- &-6.1\% &-6.7\% &-10.4\% &-4.9\% \\
Go &\cellcolor[HTML]{d9ead3}1.8\% &- &-6.1\% &-7.3\% &-6.7\% &\cellcolor[HTML]{d9ead3}0.6\% &- &-5.5\% &-12.2\% &-9.1\% \\
Java &\cellcolor[HTML]{d9ead3}1.2\% &-18.3\% &- &-11.0\% &-1.8\% &-4.3\% &-21.3\% &- &-11.0\% &-5.5\% \\
Python &\cellcolor[HTML]{d9ead3}2.4\% &0.0\% &-7.3\% &- &-2.4\% &-5.5\% &\cellcolor[HTML]{d9ead3}3.7\% &-3.7\% &- &-1.8\% \\
Rust &\cellcolor[HTML]{d9ead3}2.4\% &-9.8\% &-7.9\% &-4.3\% &- &-3.7\% &-5.5\% &-8.5\% &-9.1\% &- \\
Natural language &\cellcolor[HTML]{d9ead3}0.6\% &-7.3\% &-7.3\% &-3.0\% &-4.9\% &\cellcolor[HTML]{d9ead3}0.6\% &\cellcolor[HTML]{d9ead3}3.0\% &0.0\% &-12.2\% &-4.9\% \\
Pseudo code &\cellcolor[HTML]{d9ead3}1.8\% &-1.2\% &-11.0\% &-11.0\% &-0.6\% &-6.7\% &-7.3\% &-25.0\% &-10.4\% &-9.8\% \\
Direct generation &\cellcolor[HTML]{fff2cc}36.6\% &\cellcolor[HTML]{fff2cc}40.9\% &\cellcolor[HTML]{fff2cc}56.7\% &\cellcolor[HTML]{fff2cc}60.4\% &\cellcolor[HTML]{fff2cc}27.4\% &\cellcolor[HTML]{fff2cc}42.7\% &\cellcolor[HTML]{fff2cc}40.2\% &\cellcolor[HTML]{fff2cc}56.1\% &\cellcolor[HTML]{fff2cc}62.8\% &\cellcolor[HTML]{fff2cc}28.0\% \\\midrule
&\multicolumn{5}{c|}{Phi-3-mini-128k-instuct} & & & & & \\\cmidrule{2-6}
&Cpp &Go &Java &Python &Rust & & & & & \\\cmidrule{1-6}
Cpp &- &-6.7\% &-7.3\% &-9.8\% &\cellcolor[HTML]{d9ead3}2.4\% & & & & & \\
Go &-7.9\% &- &-3.7\% &-6.1\% &0.0\% & & & & & \\
Java &-8.5\% &-1.8\% &- &-4.3\% &\cellcolor[HTML]{d9ead3}4.9\% & & & & & \\
Python &-1.8\% &\cellcolor[HTML]{d9ead3}0.6\% &-1.8\% &- &\cellcolor[HTML]{d9ead3}0.6\% & & & & & \\
Rust &-7.9\% &-1.8\% &-11.6\% &-6.1\% &- & & & & & \\
Natural language &-4.3\% &-3.0\% &\cellcolor[HTML]{d9ead3}7.9\% &\cellcolor[HTML]{d9ead3}1.8\% &\cellcolor[HTML]{d9ead3}12.8\% & & & & & \\
Pseudo code &-3.0\% &\cellcolor[HTML]{d9ead3}9.1\% &-1.2\% &-6.1\% &\cellcolor[HTML]{d9ead3}3.0\% & & & & & \\
Direct generation &\cellcolor[HTML]{fff2cc}40.24\% &\cellcolor[HTML]{fff2cc}23.78\% &\cellcolor[HTML]{fff2cc}45.12\% &\cellcolor[HTML]{fff2cc}56.71\% &\cellcolor[HTML]{fff2cc}3.05\% & & & & & \\\cmidrule{1-6}
\end{tabular} 
\end{adjustbox}
\end{table}

\subsubsection{Deepseek, Stable and Phi-3-mini}
We present the experimental results for Deepseek, Stable, and Phi-3-mini in Table~\ref{tab:small3}, respectively. For these smaller-sized models, the impact of intermediate languages is not significant. In the case of the Deepseek-1.3b model, while we observe some improvements in Cpp, Go, and Java, these improvements are not substantial compared to direct generation results. Pseudo code shows the worst performance, with significant degradation observed for all languages. Natural language yields better or comparable results compared to direct generation but also experiences degradation for Python and Rust.
For the Stable-3b model, while using any intermediate language results in better performance, no improvement is observed in other programming languages. Natural language and pseudo code yield similar performance, experiencing significant degradation in certain cases. Another notable observation is that using Java as an intermediate step in Go generation results in around a 20\% performance decrease for both models.

For the Phi-3-mini model, Rust benefits the most from using an intermediate generation, and most intermediate languages help to improve generation. Specifically, adding the intermediate step of generating an intermediate solution boosts the correctness from 3\% to 15\%. Similarly, a 7.9\% performance gain is found in natural language-java generation compared to direct generation. Pseudo-code is a better intermediate language for Go, which demonstrates an improvement of 9.2\%.

\textbf{Key findings}:
The positive impact of intermediate language is not significant. Further, in some cases, using the intermediate step results in performance drop. Intermediate step demonstrate the most help for language which the model have little knowledge.

\begin{table}[!htp]\centering
\caption{Average pass rate of final generation across models greater than 10b parameters. Natural language outperforms other intermediate representations and direct generations.}\label{tab:avgsub}
\begin{tabular}{l|rrrrrr}\toprule
&Cpp &Go &Java &Python &Rust \\\midrule
Cpp & - &42.9\% &52.7\% &56.7\% &31.4\% \\
Go &42.8\% & -&54.2\% &55.5\% &31.7\% \\
Java &43.2\% &44.5\% & -&56.8\% &31.6\% \\
Python &44.8\% &43.4\% &52.9\% & -&29.4\% \\
Rust &44.1\% &42.8\% &54.1\% &57.3\% & -\\
Natural language &\textbf{47.9\%} &\textbf{46.1\%} &\textbf{56.7\%} &\textbf{59.2\%} &\textbf{33.8\%} \\
Pseudo code &41.1\% &44.2\% &55.9\% &57.1\% &29.2\% \\
Direct generation &46.4\% &45.7\% &54.0\% &58.6\% &32.6\% \\
\bottomrule
\end{tabular}
\end{table}

\subsubsection{Discussion}
Table~\ref{tab:avgsub} compares the average pass rates for using different intermediate languages across models larger than 10b parameters, including Mixtral 8x7b, CodeLlama13b, 34b, 70b, and GPT family. Natural language as an intermediate step consistently leads to the highest success rates across all target languages, outperforming both direct generation and other intermediate representations. The effectiveness of programming languages as intermediaries varies. For instance, Python appears to be a relatively effective intermediate language for Cpp and Rust generation. Pseudo-code shows mixed results, sometimes performing worse than direct generation (\emph{e.g.}, for Cpp and Rust). Rust generation consistently has the lowest pass rates regardless of the intermediate representation used, suggesting it is the most challenging language for these models. The improvement from using intermediate steps is most pronounced for Go and Java generation.

\noindent\fbox{\begin{minipage}{\dimexpr\textwidth-2\fboxsep-2\fboxrule\relax}

\textbf{Answers for RQ1 and RQ2}:
The impact of using intermediate representations varies across models and languages. Generally, larger models tend to benefit more from intermediate-target generations, as evidenced by the overall positive results for these high-parameter models. No single programming language universally enhances all target generations. However, natural language consistently emerges as the most effective intermediate representation across all target languages. This suggests that breaking down the problem into a natural language description before code generation is a promising approach for improving code generation accuracy.
\end{minipage}}

\subsection{Correlations between direct generations and intermediate-target generations}
In the third experiment, we provide the ground truth~(GT) to the task in the intermediate programming language~(X) and prompt the model to generate the solution in the target language~(Y). The results are available in Appendix Table~\ref{tab:GTresults}. 

We calculate the per-case correlation between direct generation in the intermediate language and target language generation (X vs. X-Y). Given the inherent difficulty in the problem set—for easier problems, models are likely to generate correct solutions in any language—we use the correlation between direct generations in the intermediate and target languages (X vs. Y) as a baseline to measure this difficulty. We then compare the X vs. X-Y correlation with the baseline correlation to understand the effect of using the intermediate language.

The complete results are shown in Appendix Table~\ref{tab:corr}. Here, we present the results for GPT3, CodeLlama34b, and CodeLlama70b, shown in Table~\ref{tab:xyx}. The first column lists the models and intermediate languages, while the subsequent columns represent target languages. For each intermediate-target pair, the table presents the X vs. X-Y correlation under the "Intermediate" column and the X vs. Y correlation under the "Direct" column.

Our analysis reveals two key findings. Firstly, the correlation between the correctness of intermediate direct generation and intermediate-target language generation is weak, comparable to the baseline correlation. The pass rate for XGT-Y (where GT represents ground truth) is significantly higher than the X-Y generation pass rate, particularly in GPT models. For instance, GPT models show an average 21.6\% improvement in pass rate when using ground truth as an intermediate step. In contrast, CodeLlama-34b and CodeLlama-70b show only 9.26\% and 3.5\% improvements, respectively.
This suggests that the quality of intermediate answers is less critical in CodeLlama models, whereas in GPT models, the correctness of intermediate solutions has a greater impact on the final generation. 

\vspace{-3mm}
\begin{table}[!htp]\centering
\caption{Intermediate (X) vs intermediate-target (X-Y) and intermediate (X) vs direct(Y) correlation results for GPT35, CodeLlama34b and CodeLlama70b. For all languages and models, X vs. X-Y correlations are weak, similar to X vs. Y correlations.}\label{tab:xyx}
\begin{adjustbox}{width=\linewidth}
\begin{tabular}{l|cc|cc|cc|cc|ccc}\toprule
&\multicolumn{2}{c}{Cpp} &\multicolumn{2}{c}{Go} &\multicolumn{2}{c}{Java} &\multicolumn{2}{c}{Python} &\multicolumn{2}{c}{Rust} \\\midrule
GPT35 &Intermediate &Direct &Intermediate &Direct &Intermediate &Direct &Intermediate &Direct &Intermediate &Direct \\\midrule
Cpp &N/A &1.00 &0.39 &0.38 &0.46 &0.37 &0.50 &0.41 &0.35 &0.38 \\
Go &0.29 &0.38 &N/A &1.00 &0.47 &0.36 &0.35 &0.43 &0.37 &0.37 \\
Java &0.30 &0.37 &0.24 &0.36 &N/A &1.00 &0.32 &0.36 &0.33 &0.39 \\
Python &0.39 &0.41 &0.41 &0.43 &0.44 &0.36 &N/A &1.00 &0.37 &0.31 \\
Rust &0.32 &0.38 &0.11 &0.37 &0.34 &0.39 &0.29 &0.31 &N/A &1.00 \\\midrule
\makecell{CodeLlama-34b} &Intermediate &Direct &Intermediate &Direct &Intermediate &Direct &Intermediate &Direct &Intermediate &Direct \\\midrule
Cpp &N/A &1.00 &0.40 &0.30 &0.59 &0.47 &0.54 &0.45 &0.37 &0.47 \\
Go &0.44 &0.30 &N/A &1.00 &0.48 &0.32 &0.48 &0.35 &0.30 &0.38 \\
Java &0.52 &0.47 &0.41 &0.32 &N/A &1.00 &0.52 &0.42 &0.48 &0.35 \\
Python &0.49 &0.45 &0.26 &0.35 &0.55 &0.42 &N/A &1.00 &0.35 &0.30 \\
Rust &0.39 &0.47 &0.30 &0.38 &0.42 &0.35 &0.39 &0.30 &N/A &1.00 \\\midrule
\makecell{CodeLlama-70b} &Intermediate &Direct &Intermediate &Direct &Intermediate &Direct &Intermediate &Direct &Intermediate &Direct \\\midrule
Cpp &N/A &1.00 &0.35 &0.30 &0.47 &0.47 &0.48 &0.45 &0.33 &0.47 \\
Go &0.34 &0.30 &N/A &1.00 &0.29 &0.32 &0.35 &0.35 &0.25 &0.38 \\
Java &0.43 &0.47 &0.43 &0.32 &N/A &1.00 &0.48 &0.42 &0.27 &0.35 \\
Python &0.41 &0.45 &0.40 &0.35 &0.51 &0.42 &N/A &1.00 &0.15 &0.30 \\
Rust &0.42 &0.47 &0.33 &0.38 &0.41 &0.35 &0.45 &0.30 &N/A &1.00 \\
\bottomrule
\end{tabular}

\end{adjustbox}
\end{table}

\noindent\fbox{\begin{minipage}{\dimexpr\textwidth-2\fboxsep-2\fboxrule\relax}
\textbf{Answers for RQ3}:
The correlation between direct generation in the intermediate language and the subsequent intermediate-target generation is comparable to the correlation between direct generations in the two languages. This similarity suggests that the observed improvement when using intermediate steps may be attributed to the chain-of-thought effect rather than specific language transfer benefits.
\end{minipage}}

\vspace{-3mm}
\subsection{Self-correction through repetitive prompting}
In this experiment, we first prompt the model twice. The answer generated in the first round is provided as context when generating the second round. Note that we do not provide any instruction to perform self-evaluation or self-correction in the second round. 

The experiment result is shown in Table~\ref{tab:rq4}. CodeLlama and other models show mixed results with two-round prompting, with some languages benefiting and others showing decreased performance. For GPT models, especially for GPT4, we observe significant benefits of this methodology (up to 20.1\% for Rust). To further explore the effect of repetitive prompting on GPT models, we increase the time of prompting to three and we present the results in Table~\ref{tab:rq4k3}. For Cpp and Go, performance decreases compared to both two-round and single-round prompting. For Java, Python, and Rust, there are still performance gains compared to the initial generation, but these improvements are not consistently better than two-round prompting.

These results suggest that while repetitive prompting can lead to performance improvements, especially for GPT models, the relationship between the number of prompting rounds and performance is not strictly linear. The effectiveness appears to vary based on the specific model and target language.

\noindent\fbox{\begin{minipage}{\dimexpr\textwidth-2\fboxsep-2\fboxrule\relax}
\textbf{Answers for RQ4}:
Repetitive prompting can improve model performance, particularly for GPT models. However, the optimal number of prompting rounds varies by model and target language, with diminishing or negative returns beyond a certain point. This suggests that while self-correction through repetition is possible, it is not uniformly beneficial across all scenarios.
\end{minipage}}

\begin{table}[!htp]\centering
\caption{Comparision of results between repeated generation (k=2) and direct generations. Significant improvements are observed with GPT models.}\label{tab:rq4}
\begin{tabular}{l|rrrrrr}\toprule
&Cpp &Go &Java &Python &Rust \\\midrule
CodeLlama-7b & -1.80\% &\cellcolor[HTML]{d9ead3}2.50\% & -4.30\% & -1.80\% & -4.90\% \\
CodeLlama-13b & -0.90\% &\cellcolor[HTML]{d9ead3}0.60\% & -1.20\% &\cellcolor[HTML]{d9ead3}1.30\% & -5.50\% \\
CodeLlama-34b & -1.80\% &\cellcolor[HTML]{d9ead3}0.00\% &\cellcolor[HTML]{d9ead3}4.90\% & -1.20\% &\cellcolor[HTML]{d9ead3}8.60\% \\
CodeLlama-70b &\cellcolor[HTML]{d9ead3}2.50\% & -3.60\% & -1.70\% &\cellcolor[HTML]{d9ead3}0.00\% & -3.60\% \\
GPT3.5 &\cellcolor[HTML]{d9ead3}4.00\% &\cellcolor[HTML]{d9ead3}0.70\% &\cellcolor[HTML]{d9ead3}\textbf{11.20\%} & -1.60\% &\cellcolor[HTML]{d9ead3}6.00\% \\
GPT4 &\cellcolor[HTML]{d9ead3}\textbf{10.30\%} &\cellcolor[HTML]{d9ead3}\textbf{5.50\%} &\cellcolor[HTML]{d9ead3}0.60\% &\cellcolor[HTML]{d9ead3}\textbf{1.90\%} &\cellcolor[HTML]{d9ead3}\textbf{20.10\%} \\
Deepseek1.3b & -2.50\% & -4.20\% &\cellcolor[HTML]{d9ead3}0.00\% & -4.90\% & -9.70\% \\
Stable-3b &\cellcolor[HTML]{d9ead3}5.50\% & -3.70\% & -3.30\% &\cellcolor[HTML]{d9ead3}1.80\% & -2.40\% \\
Phi-3-mini & -3.00\% &\cellcolor[HTML]{d9ead3}1.20\% &\cellcolor[HTML]{d9ead3}1.20\% & -3.00\% &\cellcolor[HTML]{d9ead3}6.10\% \\
\bottomrule
\end{tabular}

\end{table}

\begin{table}[!htp]\centering
\caption{Comparison of results between repeated generation (k=3) and direct generation of GPT models. While improvements are observed in Java and Python, pass rates decline for C++ and Go.}\label{tab:rq4k3}
\begin{tabular}{l|rrrrrr}\toprule
&Cpp &Go &Java &Python &Rust \\\midrule
GPT3.5 & -4.56\% & -13.91\% &\cellcolor[HTML]{d9ead3}10.02\% &\cellcolor[HTML]{d9ead3}1.45\% &\cellcolor[HTML]{d9ead3}5.44\% \\
GPT4 & -1.85\% & -4.89\% &\cellcolor[HTML]{d9ead3}14.01\% &\cellcolor[HTML]{d9ead3}9.79\% &\cellcolor[HTML]{d9ead3}12.78\% \\
\bottomrule
\end{tabular}
\end{table}

\section{Related Work}\label{sec:relatedwork}
In-context learning has gained significant attention for its ability to enhance large language models (LLMs) in code generation. Yuan et al.~\cite{yuan2023evaluating} performed experiments with instruction-tuned LLMs under various settings and concluded that providing few-shot examples improves the performance of LLMs in code generation. Similarly, Li et al. ~\cite{li2023large} constructed a novel dataset for benchmarking example selections when using LLMs for few-shot generation. Zhong et al. ~\cite{zhong2024can} discussed the challenges of using LLMs for code generation, emphasizing the importance of example quality to ensure code robustness and reliability. Agent-like behaviors, such as allowing LLMs to ask for clarifications (Mu et al. ~\cite{mu2023clarifygpt}), self-plan (Jiang et al.~\cite{jiang2023self}), or iteratively debug to improve the code (Jiang et al. ~\cite{jiang2023selfevolve}), have also been widely adopted. Additionally, various reasoning-based prompting techniques designed for code generation have been proposed, including SCoT ~\cite{li2023structured}, Think Dot-by-Dot ~\cite{pfau2024let}, and ChainCoder ~\cite{zheng2023outline}. The use of external tools has also been explored, as demonstrated in the work by Gao et al.~\cite{gao2023pal}.

Studying the performance of large language models (LLMs) across different programming languages has recently gained significant attention. Buscemi ~\cite{buscemi2023comparative} conducted an empirical analysis of ChatGPT-3.5's capabilities for generating code in ten different programming languages and discovered several weaknesses, including poor performance in less popular languages. Other similar studies have introduced unique methodologies to construct multi-language programming datasets for benchmarking and evaluation~\cite{athiwaratkun2022multi, cassano2023multipl,ding2024crosscodeeval}, which have further advanced the field. Using first-hand results from benchmark evaluations, Paul et al. ~\cite{paul2024ircoder} developed IRCoder to improve LLM code generation for various programming languages by adopting intermediate representations. Besides code generation, Joshi et al.~\cite{joshi2023repair} investigated the performance of LLMs in program repair across different programming languages and demonstrated that LLMs outperform other monolingual repair tools. Code translation, a subset of LLM code generation, has also attracted considerable interest. Pan et al.~\cite{pan2024lost} examined the quality of LLM code translation and identified weaknesses related to bug generation during the process. Macedo et al. ~\cite{macedo2024exploring} studied program translation capabilities across multiple LLMs and emphasized the importance of output format and code extraction in the reliability of LLMs. Various approaches to improving LLM code translation performance have been proposed, including using compiler representations as context~\cite{szafraniec2022code} and leveraging static analysis tools~\cite{ibrahimzada2024program}.

\section{Conclusion}\label{sec:conclusion}
Inspired by using intermediate steps to improve the performance of LLM in reasoning tasks, in this study, we systematically evaluated the use of intermediate representation, including programming languages, pseudo code, and natural language solution sketch to improve the LLM code generation process and propose the intermediate-prompting method. Through experiments with different families of models, we find that different models exhibit different behaviors, and generally, the use of intermediate representation is more effective in larger size of models. Further, we observe that natural language is a universally good intermediate step; however, there is no such formal language. Through investigation of the correlations between direct generations and intermediate-step generations, we believe that the observed improvement can be attributed to the effect of COT. The correctness of intermediate solutions demonstrates more impact on the GPT model than other models. Further, repetitive prompting also helps GPT models to generate better code.

\bibliographystyle{unsrt}  
\bibliography{refs}  

\appendix
\clearpage
\newpage
\section{Appendix}
\begin{table}[!htp]\centering
\caption{Results for ground truth-target generation.}\label{tab:GTresults}
\begin{adjustbox}{width=\linewidth} 
\begin{tabular}{l|ccccc|ccccc}\toprule
&\multicolumn{5}{c}{GPT35} &\multicolumn{5}{c}{Mixtral 8x7b} \\\cmidrule{2-11}
&Cpp &Go &Java &Python &Rust &Cpp &Go &Java &Python &Rust \\\midrule
Cpp &N/A &74.39\% &71.95\% &85.98\% &71.34\% &N/A &57.93\% &73.17\% &66.46\% &48.78\% \\
Go &69.51\% &N/A &71.34\% &89.63\% &71.95\% &60.98\% &N/A &76.83\% &51.83\% &43.29\% \\
Java &74.39\% &72.56\% &N/A &85.98\% &65.85\% &60.37\% &62.20\% &N/A &78.66\% &56.71\% \\
Python &68.29\% &72.56\% &67.68\% &N/A &68.29\% &56.10\% &60.37\% &71.34\% &N/A &42.07\% \\
Rust &73.17\% &65.85\% &73.78\% &85.37\% &N/A &59.15\% &62.80\% &70.12\% &70.12\% &N/A \\\midrule
&\multicolumn{5}{c}{CodeLlama-7b} &\multicolumn{5}{c}{CodeLlama-13b} \\\cmidrule{2-11}
&Cpp &Go &Java &Python &Rust &Cpp &Go &Java &Python &Rust \\\cmidrule{1-11}
Cpp &N/A &44.51\% &57.93\% &54.27\% &25.00\% &N/A &42.68\% &57.93\% &47.56\% &30.49\% \\
Go &47.56\% &N/A &54.88\% &62.80\% &27.44\% &48.78\% &N/A &57.32\% &58.54\% &32.32\% \\
Java &48.17\% &43.90\% &N/A &62.20\% &25.00\% &47.56\% &46.34\% &N/A &62.20\% &34.76\% \\
Python &44.51\% &39.02\% &47.56\% &N/A &25.00\% &45.12\% &41.46\% &51.22\% &N/A &23.78\% \\
Rust &39.63\% &35.37\% &42.68\% &51.83\% &N/A &48.17\% &36.59\% &53.05\% &45.73\% &N/A \\\midrule
&\multicolumn{5}{c}{CodeLlama-34b} &\multicolumn{5}{c}{CodeLlama-70b} \\
&Cpp &Go &Java &Python &Rust &Cpp &Go &Java &Python &Rust \\\cmidrule{1-11}
Cpp &N/A &39.02\% &53.05\% &58.54\% &31.10\% &N/A &54.88\% &67.07\% &67.68\% &32.32\% \\
Go &48.78\% &N/A &58.54\% &62.20\% &31.10\% &54.88\% &N/A &70.12\% &68.90\% &29.88\% \\
Java &52.44\% &44.51\% &N/A &60.98\% &34.15\% &54.88\% &56.71\% &N/A &67.07\% &30.49\% \\
Python &48.78\% &42.07\% &53.05\% &N/A &32.93\% &56.10\% &53.66\% &72.56\% &N/A &31.10\% \\
Rust &46.95\% &43.90\% &52.44\% &53.05\% &N/A &56.10\% &51.22\% &68.90\% &65.85\% &N/A \\\midrule
&\multicolumn{5}{c}{Deepseek-1.3b} &\multicolumn{5}{c}{Stable-3b} \\\cmidrule{2-11}
&Cpp &Go &Java &Python &Rust &Cpp &Go &Java &Python &Rust \\\cmidrule{1-11}
Cpp &N/A &44.51\% &65.24\% &65.24\% &28.05\% &N/A &38.41\% &49.39\% &55.49\% &22.56\% \\
Go &51.22\% &N/A &65.85\% &70.73\% &28.66\% &36.59\% &N/A &48.17\% &35.98\% &20.12\% \\
Java &59.76\% &48.17\% &N/A &73.78\% &29.88\% &39.02\% &41.46\% &N/A &57.32\% &24.39\% \\
Python &51.83\% &43.29\% &66.46\% &N/A &25.61\% &34.76\% &38.41\% &54.88\% &N/A &23.17\% \\
Rust &57.32\% &39.02\% &62.80\% &60.37\% &N/A &37.80\% &35.98\% &45.73\% &57.93\% &N/A \\\midrule
&\multicolumn{5}{c}{Phi-3-mini-128k-instruct} & & & &  \\\cmidrule{2-6}
Cpp &Go &Java &Python &Rust & & & & & &\\\cmidrule{1-6}
Cpp &N/A &32.93\% &54.27\% &62.20\% &11.59\% & & & & &\\
Go &39.02\% &N/A &25.00\% &65.24\% &13.41\% & & & & &\\
Java &37.80\% &19.51\% &N/A &50.61\% &15.85\% & & & & &\\
Python &43.29\% &29.27\% &40.24\% &N/A &11.59\% & & & & &\\
Rust &39.02\% &26.22\% &43.29\% &65.85\% &N/A & & & & &\\\cmidrule{1-6}
\end{tabular}
\end{adjustbox}
\end{table}

\begin{table}[!htp]\centering
\caption{Intermediate (X) vs intermediate-target (X-Y) and intermediate (X) vs direct(Y) correlation results.}\label{tab:corr}
\begin{adjustbox}{width=\linewidth} 

\begin{tabular}{l|cc|cc|cc|cc|ccc}\toprule
&\multicolumn{2}{c}{Cpp} &\multicolumn{2}{c}{Go} &\multicolumn{2}{c}{Java} &\multicolumn{2}{c}{Python} &\multicolumn{2}{c}{Rust} \\\midrule
GPT35 &Intermediate &Direct &Intermediate &Direct &Intermediate &Direct &Intermediate &Direct &Intermediate &Direct \\\midrule
Cpp &N/A &1.00 &0.39 &0.38 &0.46 &0.37 &0.50 &0.41 &0.35 &0.38 \\
Go &0.29 &0.38 &N/A &1.00 &0.47 &0.36 &0.35 &0.43 &0.37 &0.37 \\
Java &0.30 &0.37 &0.24 &0.36 &N/A &1.00 &0.32 &0.36 &0.33 &0.39 \\
Python &0.39 &0.41 &0.41 &0.43 &0.44 &0.36 &N/A &1.00 &0.37 &0.31 \\
Rust &0.32 &0.38 &0.11 &0.37 &0.34 &0.39 &0.29 &0.31 &N/A &1.00 \\\midrule
GPT4 &Intermediate &Direct &Intermediate &Direct &Intermediate &Direct &Intermediate &Direct &Intermediate &Direct \\\midrule
Cpp &N/A &1.00 &0.14 &0.10 &0.30 &0.25 &0.32 &0.30 &0.34 &0.15 \\
Go &0.10 &0.10 &N/A &1.00 &0.24 &0.23 &0.18 &0.25 &0.16 &0.24 \\
Java &0.27 &0.25 &0.24 &0.23 &N/A &1.00 &0.30 &0.36 &0.32 &0.27 \\
Python &0.29 &0.30 &0.00 &0.25 &0.37 &0.36 &N/A &1.00 &0.10 &0.32 \\
Rust &0.17 &0.15 &0.11 &0.24 &0.20 &0.27 &0.32 &0.32 &N/A &1.00 \\\midrule
CodeLlama7b &Intermediate &Direct &Intermediate &Direct &Intermediate &Direct &Intermediate &Direct &Intermediate &Direct \\\midrule
Cpp &N/A &1.00 &0.49 &0.43 &0.57 &0.64 &0.62 &0.59 &0.47 &0.47 \\
Go &0.49 &0.43 &N/A &1.00 &0.60 &0.55 &0.50 &0.44 &0.38 &0.34 \\
Java &0.58 &0.64 &0.45 &0.55 &N/A &1.00 &0.58 &0.60 &0.41 &0.42 \\
Python &0.53 &0.59 &0.41 &0.44 &0.61 &0.60 &N/A &1.00 &0.46 &0.48 \\
Rust &0.46 &0.47 &0.41 &0.34 &0.39 &0.42 &0.41 &0.48 &N/A &1.00 \\\midrule
CodeLlama-34b &Intermediate &Direct &Intermediate &Direct &Intermediate &Direct &Intermediate &Direct &Intermediate &Direct \\\midrule
Cpp &N/A &1.00 &0.40 &0.30 &0.59 &0.47 &0.54 &0.45 &0.37 &0.47 \\
Go &0.44 &0.30 &N/A &1.00 &0.48 &0.32 &0.48 &0.35 &0.30 &0.38 \\
Java &0.52 &0.47 &0.41 &0.32 &N/A &1.00 &0.52 &0.42 &0.48 &0.35 \\
Python &0.49 &0.45 &0.26 &0.35 &0.55 &0.42 &N/A &1.00 &0.35 &0.30 \\
Rust &0.39 &0.47 &0.30 &0.38 &0.42 &0.35 &0.39 &0.30 &N/A &1.00 \\\midrule
CodeLlama-70b &Intermediate &Direct &Intermediate &Direct &Intermediate &Direct &Intermediate &Direct &Intermediate &Direct \\\midrule
Cpp &N/A &1.00 &0.35 &0.30 &0.47 &0.47 &0.48 &0.45 &0.33 &0.47 \\
Go &0.34 &0.30 &N/A &1.00 &0.29 &0.32 &0.35 &0.35 &0.25 &0.38 \\
Java &0.43 &0.47 &0.43 &0.32 &N/A &1.00 &0.48 &0.42 &0.27 &0.35 \\
Python &0.41 &0.45 &0.40 &0.35 &0.51 &0.42 &N/A &1.00 &0.15 &0.30 \\
Rust &0.42 &0.47 &0.33 &0.38 &0.41 &0.35 &0.45 &0.30 &N/A &1.00 \\\midrule
Mistral-7b &Intermediate &Direct &Intermediate &Direct &Intermediate &Direct &Intermediate &Direct &Intermediate &Direct \\\midrule
Cpp &N/A &1.00 &-0.02 &0.07 &0.06 &0.05 &-0.09 &-0.04 &0.01 &-0.03 \\
Go &0.11 &0.07 &N/A &1.00 &0.09 &0.11 &-0.01 &0.01 &0.03 &-0.02 \\
Java &0.01 &0.05 &0.05 &0.11 &N/A &1.00 &0.07 &0.05 &-0.06 &-0.05 \\
Python &-0.03 &-0.04 &0.03 &0.01 &0.03 &0.05 &N/A &1.00 &0.03 &0.06 \\
Rust &-0.00 &-0.03 &-0.01 &-0.02 &-0.03 &-0.05 &0.09 &0.06 &N/A &1.00 \\\midrule
Mixtral8x7b &Intermediate &Direct &Intermediate &Direct &Intermediate &Direct &Intermediate &Direct &Intermediate &Direct \\\midrule
Cpp &N/A &1.00 &0.44 &0.48 &0.46 &0.46 &0.35 &0.36 &0.51 &0.39 \\
Go &0.46 &0.48 &N/A &1.00 &0.41 &0.53 &0.32 &0.40 &0.35 &0.25 \\
Java &0.46 &0.46 &0.58 &0.53 &N/A &1.00 &0.56 &0.44 &0.41 &0.31 \\
Python &0.33 &0.36 &0.32 &0.40 &0.51 &0.44 &N/A &1.00 &0.38 &0.34 \\
Rust &0.38 &0.39 &0.33 &0.25 &0.43 &0.31 &0.27 &0.34 &N/A &1.00 \\\midrule
Phi-mini-3 &Intermediate &Direct &Intermediate &Direct &Intermediate &Direct &Intermediate &Direct &Intermediate &Direct \\\midrule
Cpp &N/A &1.00 &0.22 &0.42 &0.49 &0.51 &0.50 &0.44 &0.24 &0.14 \\
Go &0.35 &0.42 &N/A &1.00 &0.29 &0.30 &0.24 &0.31 &0.32 &0.15 \\
Java &0.41 &0.51 &0.38 &0.30 & &1.00 &0.40 &0.54 &0.28 &0.20 \\
Python &0.51 &0.44 &0.30 &0.31 &0.49 &0.54 &N/A &1.00 &0.17 &0.15 \\
Rust &0.10 &0.14 &0.08 &0.15 &0.02 &0.20 &0.18 &0.15 &N/A &1.00 \\\midrule
Stable-3b &Intermediate &Direct &Intermediate &Direct &Intermediate &Direct &Intermediate &Direct &Intermediate &Direct \\\midrule
Cpp &N/A &1.00 &0.45 &0.32 &0.52 &0.46 &0.44 &0.38 &0.25 &0.07 \\
Go &0.47 &0.32 &N/A &1.00 &0.37 &0.35 &0.36 &0.27 &0.19 &0.18 \\
Java &0.35 &0.46 &0.21 &0.35 &N/A &1.00 &0.49 &0.47 &0.27 &0.21 \\
Python &0.42 &0.38 &0.42 &0.27 &0.48 &0.47 &N/A &1.00 &0.30 &0.30 \\
Rust &0.18 &0.07 &0.21 &0.18 &0.19 &0.21 &0.32 &0.30 &N/A &1.00 \\\midrule
Deepseek-1.3b &Intermediate &Direct &Intermediate &Direct &Intermediate &Direct &Intermediate &Direct &Intermediate &Direct \\\midrule
Cpp &N/A &1.00 &0.13 &0.45 &0.38 &0.47 &0.18 &0.43 &0.40 &0.23 \\
Go &0.39 &0.45 &N/A &1.00 &0.24 &0.40 &0.04 &0.37 &0.37 &0.18 \\
Java &0.29 &0.47 &0.08 &0.40 &N/A &1.00 &0.23 &0.41 &0.36 &0.28 \\
Python &0.41 &0.43 &0.30 &0.37 &0.38 &0.41 &N/A &1.00 &0.37 &0.37 \\
Rust &0.28 &0.23 &0.17 &0.18 &0.30 &0.28 &0.23 &0.37 &N/A &1.00 \\
\bottomrule
\end{tabular}
\end{adjustbox}
\end{table}

\end{document}